\documentclass[prd,nofootinbib,twocolumn,showpacs]{revtex4}
\usepackage{graphics} 
\usepackage{bm} 
 
\begin{document}

\title{Forecasting Cosmic Doomsday from CMB/LSS Cross-Correlations} 
 
\author{Jaume Garriga$^{1,2}$, Levon Pogosian$^2$ and Tanmay Vachaspati$^3$}

\smallskip

\affiliation{$^1$Departament de Fisica Fonamental, Universitat de Barcelona, 
Marti i Franques, 1, 08028 Barcelona, Spain. \\
\smallskip
$^2$Institute of Cosmology, Department of Physics and Astronomy, 
Tufts University, Meford, MA 02155 USA. \\
\smallskip
$^3$CERCA, Department of Physics, Case Western Reserve University, 
10900 Euclid Avenue, Cleveland, OH 44106-7079, USA.} 

\begin{abstract} 
A broad class of dark energy models, which have been proposed in attempts at 
solving the cosmological constant problems, predict a late time variation of 
the equation of state with redshift. The variation occurs as a scalar field 
picks up speed on its way to negative values of the potential. The negative 
potential energy eventually turns the expansion into contraction and the local 
universe undergoes a big crunch. In this paper we show that cross-correlations 
of the CMB anisotropy and matter distribution, in combination with other 
cosmological data, can be used to forecast the imminence of such cosmic 
doomsday.
\end{abstract}


\date{\today}

\maketitle

\section{Introduction}
\label{intro}

Observational cosmology has yielded several surprises, of which the
most perplexing is the discovery of a smooth dark energy (DE) component 
which has come to dominate the universe at recent epochs, causing cosmic 
acceleration \cite{Riess,Perlmutter}.  The nature of this component is 
still a matter of speculation, and a very important challenge for the coming 
years will be to determine its origin and physical properties. Recently, 
several groups \cite{xray,nolta,fosalba,scranton,afshordi} have reported 
positive results for the cross-correlation 
between the CMB power spectrum and that of different LSS surveys, providing 
further evidence for the existence of DE. 
In this paper, we shall try and 
illustrate how such cross-correlation may help in unveiling some of the 
properties of DE, focusing on the observational signatures of a model with 
a time dependent DE equation of state.

The simplest interpretation of the
dark energy is in terms of a cosmological constant, with equation of state 
$p_D=-\rho_D$. The cosmological constant, however, raises
two puzzles of its own.  First, there is a glaring discrepancy
between the observed dark energy density $\rho_D$ and the huge values
of the cosmological constant suggested by particle physics models.
Second, the observed $\rho_D$ is comparable to the matter density
$\rho_m$.  This is the notorious time coincidence problem: why do we
happen to live at the epoch when the dark energy starts dominating?

It has long been suggested that both puzzles may find a natural explanation 
through anthropic selection effects, in scenarios where $\rho_D$ is a random 
variable, taking different values in different parts
of the universe.  The proposed selection mechanism is very
simple
\cite{Linde84,AV95,Efstathiou95,Weinberg87,MSW,GLV,Bludman}. The growth of density 
fluctuations leading to galaxy formation effectively stops when $\rho_D$ comes to
dominate over the matter density.  In regions where $\rho_D$ is greater, it will 
dominate earlier, and thus there will be fewer galaxies (and therefore
fewer observers).  A typical observer should then expect to find herself in a
region where $\rho_D$ dominates at about the epoch of galaxy formation
(which is close to the present time).  Much larger values of $\rho_D$
would yield no galaxies at all, while much smaller values are unlikely
due to the smallness of the corresponding range of $\rho_D$, assuming
that all values of $\rho_D$ are {\it a priori} equally likely \cite{Weinberg87,GV00}. 

A simple implementation of this idea is obtained by assuming that the
dark energy is due to a scalar field $\phi$ with a very flat potential
$V(\phi)$ \cite{Linde86,GV00}.  The values of $\phi$ are randomized by
quantum fluctuations during inflation, resulting in a variation of
$\phi$ with a characteristic scale much greater than the present
Hubble radius.  Galaxy formation is possible only in regions where
$V(\phi)$ is in a narrow range near $V=0$.  One expects that the
potential in this range is well approximated by a linear function
\cite{GV00,GV03,DT03},
\begin{equation}
V(\phi)= -s\phi,
\label{linear}
\end{equation}
where $s\equiv -V'(0)$ and we have set $\phi=0$ at $V=0$.  The slope
$s$ should be sufficiently small, so that the variation of $\phi$
is not fast on the present Hubble scale.  Quantitatively, this can be
expressed as the slow roll condition,
\begin{equation}
s\lesssim 3 H_0^2 M_p,
\label{slowroll}
\end{equation} 
where $M_p=1/\sqrt{8 \pi G}$ is the reduced Planck mass, $G$ is Newton's 
gravitational constant, and $H_0$ is the present Hubble expansion rate.  

In models with a single DE field, and in the absence of ad-hoc adjustments, it has been argued in \cite{GV03} that the slow roll condition (\ref{slowroll}) is likely to be satisfied by excess, by 
many orders of magnitude, rather than marginally.  In this case, $\phi$
remains nearly constant on the Hubble scale, and the effective
equation of state for the dark energy is $w\equiv p_D/\rho_D\approx
-1$, with a very high accuracy.
However, a different situation may be expected in multi-field models, where the slope 
of the potential $s$ is itself a random variable \cite{mistake,GLV03} (the role 
of the field $\phi$ in multi-field
models is played by the variable in the direction of $\nabla
V(\phi_a)$ in the field space.) The observed value of the slope may then be 
influenced by anthropic selection.  A very large slope
would cause a big crunch much before any observers can develop. Hence, in 
cases where the prior distribution favors large $s$, the most probable values
of the slope would be the ones for which the slow roll condition (\ref{slowroll}) 
is only marginally satisfied
\cite{GLV03} (see also \cite{DT03}).  

We thus have some motivation to consider a model where the dark energy
is due to a scalar field with a linear potential (\ref{linear}) and a
slope $s$ marginally satisfying the slow roll condition
(\ref{slowroll}).  
A marginal value of the slope
implies that the big crunch is imminent in about 10 billion years from
now.  The model can therefore be called a ``doomsday model''. A salient 
feature of this scenario is that the equation of state of dark energy changes 
significantly at low redshift, when the correlation between the large scale 
structure evolution and the CMB temperature anisotropies develop. Hence, we 
may expect that the analysis of such cross correlation may reveal a time 
varying equation of state $p_D = w(z) \rho_D$, where $z$ is the redshift. 
This will be the subject of the present paper. 

Prospective constraints on cosmic doomsday based on future determinations of 
the dimming of distant supernova were discussed in Ref.\cite{KKLLS03}. The 
analysis shows that SNIa observations, in combination with CMB and weak 
lensing data, have an impressive potential for constraining the equation of 
state parameter. However, the constraints reported in \cite{KKLLS03} still 
show a considerable degeneracy amongst models with the same 
``average'' $\langle w \rangle$ 
([see Eq. (\ref{average}) below). Interestingly, as we shall see, 
the ISW-matter 
cross correlation breaks this degeneracy, offering the possibility 
of telling a true doomsday model from a model with a constant 
$w=\langle w \rangle$.

The methods presented here can obviously be used in a more general context, 
provided that there is significant evolution of $w$ at low redshifts. 
A rather common assumption in phenomenological studies of dark energy 
is to consider the simplest case of a constant $w$. This is partially 
motivated by degeneracies such as the one we just discussed above, which 
also occur in the angular spectrum of CMB anisotropies
\cite{caldwell} as well as in the linear matter power spectrum. The analysis 
of CMB/LSS cross correlations with a constant $w$ (including $w=-1$) was 
considered in Refs.~\cite{turok96,peiris00,cooray,bean}. 
Here we shall drop this assumption, since the variation of $w$
with redshift may provide a very exciting clue to the nature of dark energy, 
as discussed above.

In our calculation we shall adopt a top-down approach, starting from the 
primordial spectrum of fluctuations. This differs from previous studies
where the starting point is the present matter power 
spectrum (which is evolved backward in order to find its correlation 
with CMB). Our approach unifies the treatment of CMB and matter power 
spectra, and is more convenient for taking full account of fluctuations 
in the dark energy. Moreover, since all perturbations are evolved 
numerically with the CMBFAST code \cite{cmbfast}, 
we do not resort to the frequently used approximate analytical expressions 
for the growth function, or the also commonly used small angle 
approximation \cite{Limber}. The details of our calculation are reported 
in the Appendix.

The paper is organized as follows. In Section~\ref{evolution} we describe
the evolution of the universe according to the doomsday model. We also show 
that the corresponding CMB and present matter power spectra are 
virtually identical to those which are obtained in a model with a constant 
$w$, equal to the average $\langle w \rangle$ for the doomsday model. 
In Section~\ref{temp-matter} we study the matter/CMB temperature 
cross-correlation, 
and show that it can be used to break this degeneracy. Our conclusions 
are summarized in Section~\ref{summary}.

\section{Evolution and power spectra in the doomsday scenario}
\label{evolution}

We are interested in the late time evolution of our observable universe, and 
so we assume a background model which is homogeneous and isotropic. In addition 
to the scalar field $\phi$ with linear potential (\ref{linear}), the universe 
contains the usual radiation and matter.
The dynamics is given by the Friedman-Robertson-Walker (FRW) 
equation and the scalar field equation
\begin{equation}
H^2 \equiv \left ( \frac{\dot a}{a} \right )^2 =
              \frac{1}{3M_p^2}\left(\frac{{\dot \phi}^2}{2} - s\phi \right) + 
             \frac{\Omega_{m0} H_0^2}{a^3} \ ,
\label{frweq}
\end{equation}
\begin{equation}
{\ddot \phi} + 3 H {\dot \phi} - s = 0.
\label{phieq}
\end{equation}
Here, $\Omega_{m0}$ is the fractional energy density in matter today, and 
$H_0$ is the present Hubble rate.

In addition to the equations of motion, we need to specify
initial or boundary conditions. At early times ($t \to 0$),
we expect that the scalar field is at rest and so ${\dot \phi}=0$.
This is because, in the back of our minds, we imagine an inflationary
phase which redshifts the gradients and velocities of the
scalar field. Hence the scalar field is effectively homogeneous
and static at $t=0$. The initial value of the scale factor is
zero as in usual Friedmann-Robertson-Walker evolution. The
initial value of the scalar field is a free parameter. As mentioned in the introduction, this takes different values in distant regions of the universe, separated by distances much larger than the present Hubble radius. Finally,
we require that the present value of the total energy density
be unity. This is a boundary condition. These conditions can
be summarized as follows:
\begin{equation}
a(0)=0 \ , \ \ \phi(0) = \phi_0 \ , \ \ {\dot \phi}(0) =0 \ ,
\ \ H^2 (t_0 ) = H_0^2 \ ,
\label{initconditions}
\end{equation}
where $t_0$ is the present time, defined by the requirement
that $a(t_0 ) =1$.

The cosmological evolution following from Eqs.~(\ref{frweq}), 
(\ref{phieq}) and (\ref{initconditions}) has been studied by several 
authors \cite{KLPS02,GV03,KKLLS03,GLV03}. The main features of
the evolution are as follows:
\begin{itemize}
\item The universe starts out dominated
by matter and hence $a \sim t^{2/3}$. At the same time the
scalar field is essentially at rest. 
\item After some time, the 
matter density falls below that of the scalar field potential
energy, and the evolution becomes scalar field dominated.
Since most of the energy in the scalar field is potential
energy, we have $a \propto \exp (H t )$. 
\item As the field
slips down the potential, the potential energy changes sign once 
$\phi$ changes sign. With further slipping, a time comes when the 
total energy density is zero. This epoch marks the turning point, 
where cosmic expansion changes to contraction. From this time on, 
$H$ is given by the {\it negative} square root of the right-hand 
side of Eq.~(\ref{frweq}).
\item As the universe starts contracting, the kinetic energy of the scalar 
field comes to dominate. But there is no stopping the contracting phase, and the 
universe rapidly arrives at the big crunch. 
\end{itemize}
\begin{figure}[tbp]
\centering
\scalebox{0.4}{\includegraphics{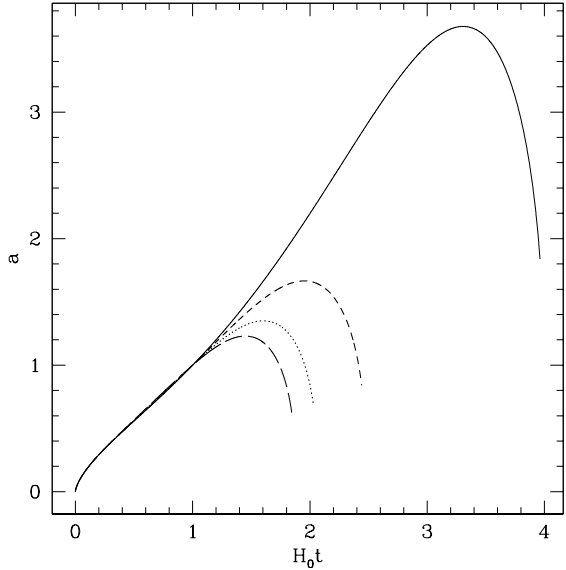}} 
\caption{Scale factor versus time for four different slopes:
$\tilde s=1$ (solid), $\tilde s=2$ (short dash), $\tilde s=3$ (dot) and 
$\tilde s=4$ (long dash)}
\label{avst} 
\end{figure}
These features are apparent in Fig.~\ref{avst}, where the
time evolution of the universe is represented for different values of the 
dimensionless slope 
\begin{equation}
\tilde s\equiv s/(\sqrt{3}M_p H_0^2).
\label{tildes}
\end{equation}
One feature of this evolution that is relevant for
observational cosmology is that the equation of state
for the scalar field changes in an unconventional manner.
Very early on, ${\dot \phi} \to 0$ and
hence $p_D \sim - \rho_D$ where $p_D$ and 
$\rho_D$ denote pressure and energy density in $\phi$.
The dark energy equation of state parameter is defined by
\begin{equation}
w \equiv \frac{p_D}{\rho_D} = 
\frac{{\dot \phi}^2/2 - s \phi}{{\dot \phi}^2/2 + s \phi} \ .
\end{equation}
Since ${\dot \phi} \simeq 0$ at early times, $w \simeq -1$. 
At later times, the
field starts to roll down the potential and hence the
kinetic energy starts to play a role in $p_D$
and $\rho_D$. This means that $w$ {\em increases}
with time. In Fig.~\ref{wvsz} we show $w$ as a function of
redshift $z$ (Figures analogous to Figs.~\ref{avst} and \ref{wvsz} can 
also be found  in \cite{KKLLS03,DT03}).

\begin{figure}[tbp]
\centering
\scalebox{0.4}{\includegraphics{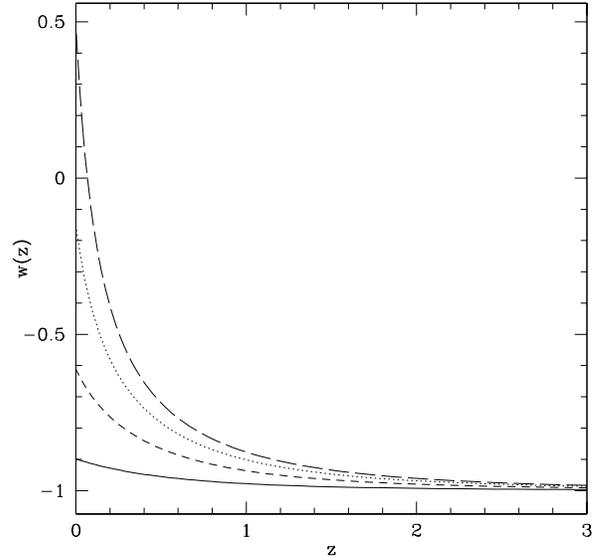}} 
\caption{$w(z)$ versus $z$ for the four different slopes
considered in Fig~\ref{avst}.}
\label{wvsz} 
\end{figure}

If we only varied $\tilde s$, keeping all other cosmological parameters
fixed, we would change the angular diameter distance to the last scattering
surface and spoil the agreement with current CMB data. Hence
it is necessary to suitably adjust cosmological parameters for different
values of $\tilde s$. The cosmological parameters in models with different 
values of $\tilde s$, presented in Figs.~\ref{avst} and \ref{wvsz}, 
depend on $\tilde s$. The simplest
way to preserve the shape of the CMB fluctuations spectra on small
and intermediate scales (where the cosmic variance is small) is to explore 
the so-called geometric degeneracy \cite{BondEfstathiou}. The shape of
the spectra depends mainly on two scales: $r_s$ -- the sound horizon at the
time of recombination, and $d_A$ -- the angular diameter distance to the
last scattering surface. Models with same values of $\Omega_M h^2$ and 
$\Omega_b h^2$ have the same $r_s$. Here $\Omega_M$ and $\Omega_b$
are the total matter and baryon density fractions today, and 
$h = H_0/(100$ km ${\rm sec}^{-1}$ ${\rm Mpc}^{-1})$.
Since the dark energy did not play a significant role at the time
of recombination, changing the value of $\tilde s$, or $w$, does not 
affect $r_s$. The main effect changes in the properties of the dark energy
have on the CMB spectra on small angular scales are due to the change in 
$d_A$, which manifests itself as a shift in the positions of the peaks in 
the angular spectra. This shift can be compensated for, without altering 
the structure of the peaks, by adjusting the value of $h$.
As a reference model we use WMAP's best fit power law $\Lambda$CDM
model \cite{wmap_spergel} with $h=0.72$, spectral index $n=0.99$, 
reionization optical depth $\tau_r=0.166$, $\Omega_b h^2=0.024$, 
$\Omega_M h^2 = 0.14$ and amplitude $A=0.86$ 
(as defined in \cite{wmap_verde}).
\begin{table}[htbp]
\vskip 0.5 truecm
\begin{tabular}{|c||c|c|c|} \hline
{\bf Model} \  & \ {\bf h } \ 
& \ {\bf $\langle w \rangle_{[0,z_{ls}]}$ } \ \\ 
\hline\hline $\tilde s=0$ \ & \ 0.72 \ &  \ -1 \ \\ 
      \hline $\tilde s=1$ \ & \ 0.69 \ & \ -0.94 \ \\ 
      \hline $\tilde s=2$ \ & \ 0.66 \ &  \ -0.81 \ \\ 
      \hline $\tilde s=3$ \ & \ 0.62 \ & \ -0.66 \ \\ 
      \hline
\end{tabular}
\caption{Models considered in the paper. $\tilde s$ is the 
dimensionless slope of the potential, defined in Eq. (\ref{tildes}). 
For each value of $\tilde s$, the dimensionless Hubble parameter 
$h$ is adjusted so that the model reproduces the CMB peak 
structure observed by WMAP. The table also shows the average 
value of $w$ for different models.}
\label{table}
\end{table}
Given a value of $\tilde s$, we vary the value of $h$, while keeping
$\tau_r$, $A$, $n$, $\Omega_M h^2$ and $\Omega_b h^2$ fixed, and find one that 
best reproduces the CMB spectra of the reference model. 

As was noted in \cite{caldwell}, except for the very large scales, 
CMB spectra are only sensitive to the averaged value of the equation of state 
of dark energy, $\langle w \rangle$, defined as
\begin{equation}
\langle w \rangle_{[0,z_{ls}]} ={ \int_{a_{ls}}^{1} d a \ \Omega_D(a) w(a) 
\over {\int_{a_{ls}}^{1} d a \ \Omega_D(a)} }  \ , 
\label{average}
\end{equation}
where the subindex ${ls}$ on $z$ and $a$ refers to the surface of last 
scattering. In Table \ref{table} we show the best fit values 
of $h$ and the corresponding
values of $\langle w \rangle_{[0,z_{ls}]}$ for several values of $\tilde s$.
Note that the uncertainty in WMAP's estimate of $h$ was about $\pm 0.05$
\cite{wmap_spergel}. The same uncertainty would apply to the best fit
values of $h$ corresponding to models with $\tilde s \ne 0$.

\begin{figure}[tbp]
\centering
\scalebox{0.4}{\includegraphics{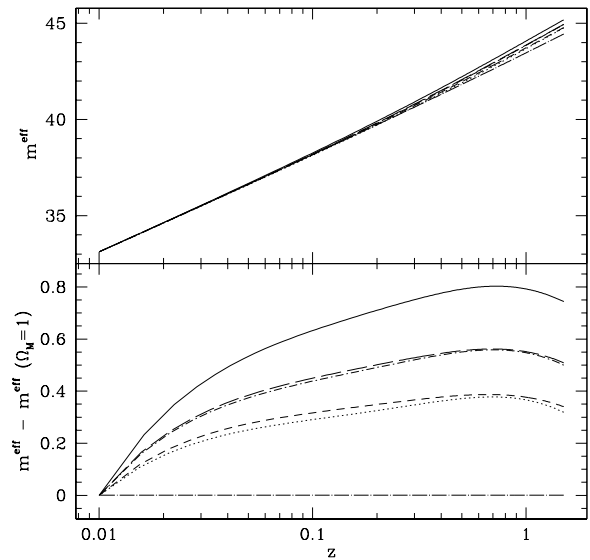}} 
\caption{ Upper panel: the luminosity, $m^{\rm eff}$, vs redshift plots 
for the $\Lambda$CDM 
(solid line), $\tilde s=2$ (dot - short dash), $\tilde s=3$ (dot), 
$w=-0.81$ (long dash), $w=-0.66$ (short dash) and the 
$\Omega_M=1$ (dot - long dash) models. 
Lower panel: differences between the $m^{\rm eff}$ for the models in the upper
panel and the $m^{eff}$ for the $\Omega_M=1$ model. Models with the same
average value of $w$ are practically indistinguishable.}
\label{lumin} 
\end{figure}

In the doomsday model, the equation of state parameter can vary significantly 
at recent redshifts. This variation is 
not necessarily constrained by existing analysis of supernova data 
\cite{knopp,dragan}, which by and large assumed a constant $w$.
In Fig.~\ref{lumin} we plot the effective luminosity $m^{\rm eff}$ 
as a function of redshift, as defined in \cite{Perlmutter}, 
for the $\tilde s=2$ and $w=-0.81$, and $\tilde s=3$ and $w=-0.66$ models.
As one can see from the Figure, the doomsday model with a given $\tilde s$
and the corresponding constant $w$ model
have almost identical predictions in the magnitude versus
redshift curves. Since constant $w$ models with $w > -0.8$
are disfavored at the $1 \sigma$ level \cite{knopp}, the value
$\tilde s =3$ which gives an average $w$ of $-0.66$
is also disfavored at the 1$\sigma$ level. However, we shall
still include it in our subsequent analysis, since it not excluded
at the $2 \sigma$ level.

\begin{figure}[tbp]
\centering
\scalebox{0.4}{\includegraphics{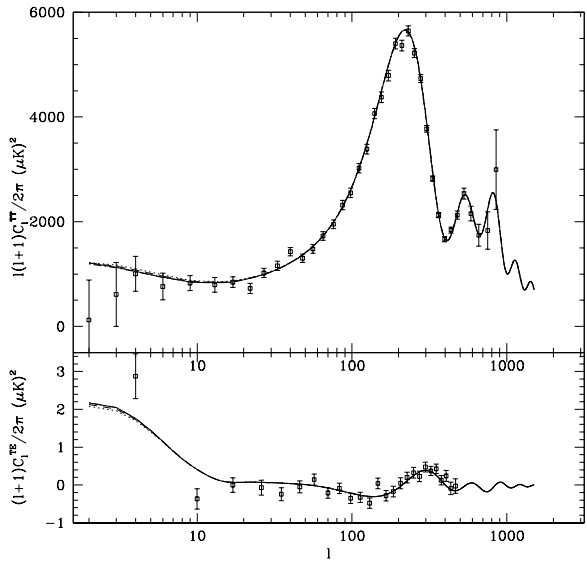}} 
\caption{TT and TE angular spectra for $\tilde s=0$ ($\Lambda$CDM) 
(solid line),
$\tilde s=1$ (dot - short dash), $\tilde s=2$ (short dash) and $\tilde s=3$ (dot) 
models together with the WMAP data points. 
The value of $h$ is chosen as indicated in Table I. Note that due to the 
geometric degeneracy, all curves look very much alike, 
the effect of a higher $\tilde s$ being undone by choosing a lower $h$.}
\label{many_s_cl} 
\end{figure}

In Fig. \ref{many_s_cl} we plot the temperature (TT) and 
temperature-polarization
cross-correlation (TE) spectra for models in Table \ref{table}.
As shown in Table I, as we increase $\tilde s$, a
smaller value of $H_0$ is needed in order to fit the CMB spectra
\footnote{In principle, one could vary other cosmological parameters as well. 
However, unless one changed the model considerably, {\it e.g.} relaxed the
assumptions of adiabaticity or scale-invariance of primordial fluctuations,
it is unlikely that one could avoid making $H_0$ small in models with
$\langle w \rangle > -1$ \cite{wmap_spergel}.}.
The value $h=-0.62 \pm 0.05$ which we
used for the $\tilde s=3$ model is somewhat lower than the currently favored 
observational $1\sigma$ region, given by $h=0.72\pm .08$ \cite{hconstraint}, 
but still marginally consistent with it.

\begin{figure}[tbp]
\centering
\scalebox{0.4}{\includegraphics{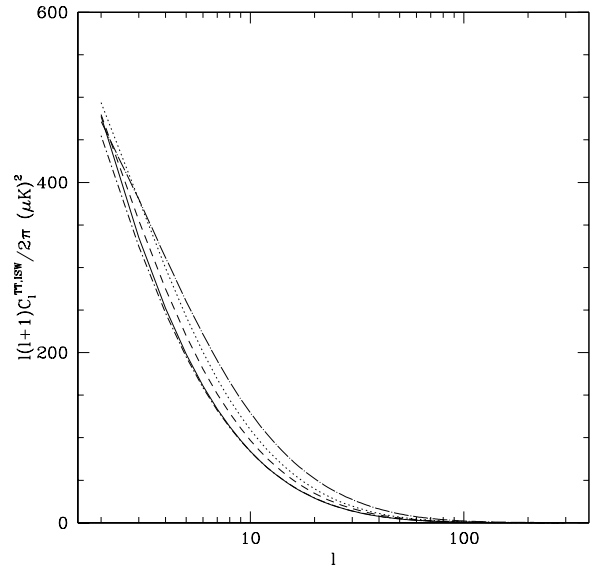}} 
\caption{The CMB temperature anisotropy angular spectrum due to the late 
ISW effect, 
$C^{TT,ISW}_\ell$,
for the five models in Fig.~\ref{many_s_cl} as well as for the model with
the constant $w=-0.66$ (dot - long dash line). As explained in 
the text, there is no suppression of the ISW effect due to a higher value 
of $w$.}
\label{isw-angle} 
\end{figure}

\begin{figure}[tbp]
\centering
\scalebox{0.4}{\includegraphics{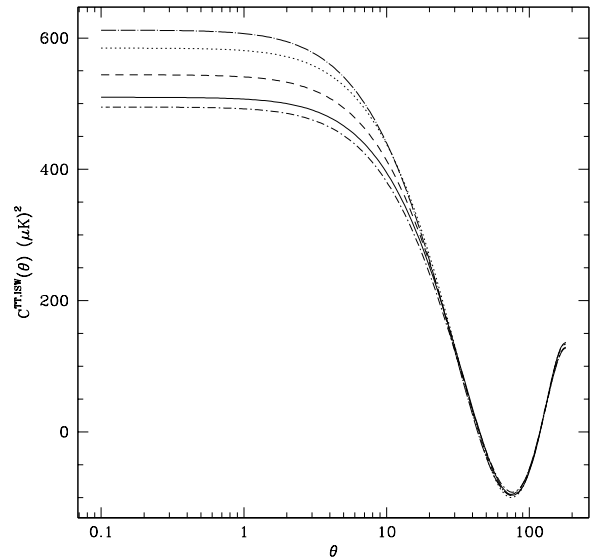}} 
\caption{The ISW sourced temperature anisotropy correlation function
$C^{TT,ISW}(\theta)$ for the models in Fig.~\ref{isw-angle}.}
\label{isw-theta} 
\end{figure}

The equation of state parameter $w(z)$ obviously has an impact on different 
observables, 
such as CMB and matter power spectra. In particular, it should affect the 
integrated Sachs-Wolfe (ISW) contributions to CMB anisotropies.
At first sight, one may think that a less negative equation of state for dark 
energy would result in a suppression 
of ISW, relative to the case of a cosmological constant. In a universe 
dominated by matter, with pressure 
$p_M=0$, there is no time dependence of the gravitational potential, and 
the frequency of a photon is only redshifted by the cosmological expansion. 
In this case, there is no late ISW effect. In the doomsday scenario, the 
dark energy equation of state is closer to that of ordinary matter, and one 
might expect that the ISW effect would be smaller than in the $w=-1$ case. 
However, the dark energy perturbations are coupled to the dark matter 
perturbations, and will also contribute to the ISW effect. The net
result is that there is no suppression of the ISW effect even for
values of $\tilde s$ corresponding to average 
$w$ as high as $-0.66$. This is illustrated in Figs.~\ref{isw-angle} 
and \ref{isw-theta}. There, we plot 
the angular spectrum $C^{TT,ISW}_\ell$ and the auto-correlation function 
$C^{TT,ISW}(\theta)$ of  temperature anisotropy due to the late ISW effect, 
defined as
\begin{eqnarray}
C^{TT,ISW}(\theta) \equiv \langle \Delta T_{ISW}(\hat n_1)  
\Delta T_{ISW}(\hat n_2)  \rangle \nonumber \\ =
\sum_{l=0}^{\infty} {2\ell+1 \over 4\pi} C^{TT,ISW}_\ell 
P_\ell({\rm cos}\ \theta).
\end{eqnarray}
Here $\Delta T_{ISW}(\hat n)$ is the ISW contribution to the temperature 
anisotropy in the direction $\hat n$ on the sky,
$\theta$ is the angle between directions $\hat n_1$ and $\hat n_2$,
and the angular brackets denote ensemble averaging (the expression
for $\Delta T_{ISW}(\hat n)$ is given in the Appendix). Note that the 
late ISW contribution of most models is in fact a bit larger than in the 
$\Lambda$CDM case.

It is quite clear from 
Fig.~\ref{many_s_cl} that CMB spectra alone are not capable of 
differentiating between models with different values of $\tilde s$ 
(including $\tilde s=0$) because of the geometric degeneracy. The effect 
of a larger $\tilde s$ can be undone with a smaller $h$. (Of course, if we 
had a stronger observational constraint on $h$, then this would result in 
stronger constraints on $\tilde s$). CMB spectra alone also cannot 
differentiate
between a model with a certain value of $\tilde s$ and a model with
the corresponding constant $w=\langle w \rangle_{[0,z_{ls}]}$ 
(see Table~\ref{table}).
\begin{figure}[tbp]
\centering
\scalebox{0.4}{\includegraphics{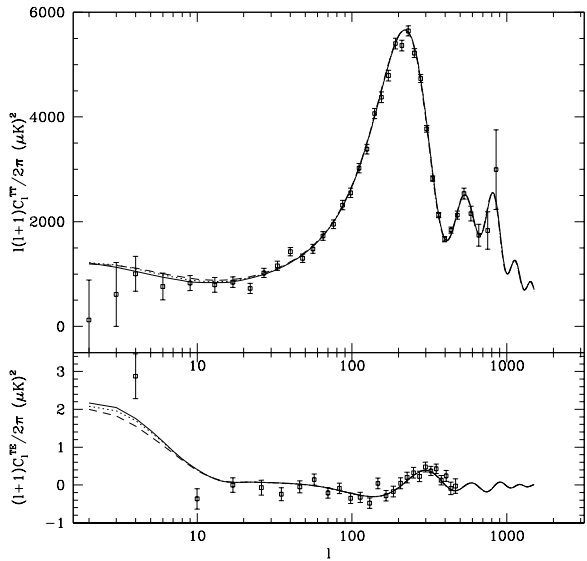}} 
\caption{$C_l$ vs $l$ for $\Lambda$CDM (solid), $s=3$ (dot) and $w=-0.66$ 
(dash) models.}
\label{clvsl} 
\end{figure} 
This is illustrated in Fig.~\ref{clvsl}, where we plot the predictions
of the $\Lambda$CDM, $w={\rm const}=-0.66$ and $\tilde s=3$ models.
\begin{figure}[tbp]
\centering
\scalebox{0.4}{\includegraphics{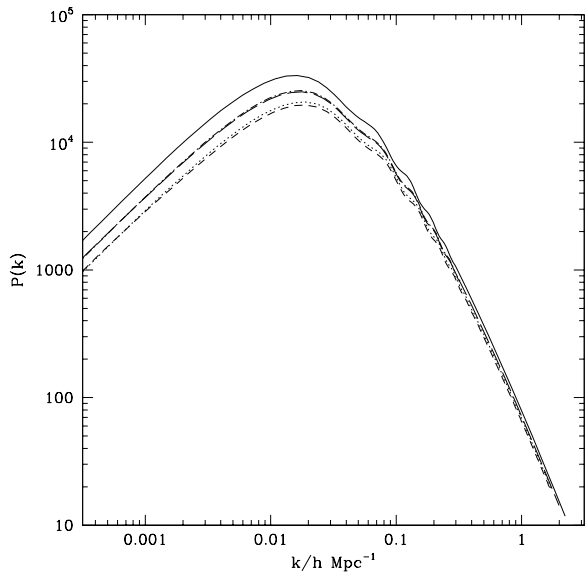}} 
\caption{The present linear power spectrum $P(k)$ vs $k$ is sensitive 
to $\langle w \rangle$, but a model with a constant $w$ has virtually the 
same $P(k)$ as a doomsday models with the same $\langle w \rangle$. Here, 
we plot $P(k)$ for the models in Fig.~\ref{clvsl}, as well as
for the $\tilde s=2$ (dot - short dash) and the $w=-0.81$ (long dash) models.}
\label{pvsk} 
\end{figure}
Ideally, some of the degeneracy can be removed by considering the matter power 
spectrum. Assuming that we had a good control over the bias, the matter 
power spectrum could in 
principle be inferred from observations. 
In Fig.~\ref{pvsk} we plot the linear matter power spectra at $z=0$ for
the $w={\rm const}=-0.66$ and $\tilde s=3$ models as, as well as for
the $\tilde s=2$ and its corresponding $w={\rm constant}=-0.81$ model.
Note that $P(k)$ differs substantially on 
large scales for $\Lambda$CDM and the models with a lower average value of $w$. 
However, there is still an impressive degeneracy
between models with the same $\langle w \rangle$: the curve corresponding 
to a constant $w=-0.66$ and the curve corresponding to the model 
with $\tilde s=3$ 
(with $\langle w \rangle =-0.66$) are almost identical, and similarly
for the $\tilde s=2$ and the $w=-0.81$ models.
In the next section we shall concern ourselves with breaking this residual 
degeneracy.

\section{ISW and temperature-matter density correlations}
\label{temp-matter}

In this section we show how the CMB/LSS cross-correlation can
be used to probe the time-dependence of the dark energy equation
of state. The cross-correlation is defined as
\begin{eqnarray}
C^{MT}(\theta) \equiv  
\langle \Delta(\hat{\bf n}_1) \delta(\hat{\bf n}_2) \rangle 
=
\sum_{l=2}^{\infty} {2\ell+1 \over 4\pi} C^{MT}_\ell P_\ell(\theta) \ ,
\end{eqnarray}
where $\Delta(\hat{\bf n}_1)$ and $\delta(\hat{\bf n}_2)$ are the CMB temperature 
anisotropy and the matter density contrast along
directions $\hat{\bf n}_1$ and $\hat{\bf n}_2$ separated by the angle $\theta$ on 
the sky\footnote{The monopole and the dipole contributions depend on the choice
of the reference frame and are not included.}. In the Appendix we show 
that the angular cross-correlation
spectrum $C^{MT}_\ell$ can be written as
\begin{equation}
C^{MT}_\ell = 4\pi {9 \over 25}\int {dk \over k} \ \Delta^2_{\cal R} \ 
T^{ISW}_\ell(k) \ M_\ell(k) \ ,
\label{crosscor}
\end{equation}
were $\Delta^2_{\cal R}$ is the primordial curvature power spectrum, as defined, 
{\it e.g.} in 
\cite{Liddle}, and $T^{ISW}_\ell(k)$ and $M_\ell(k)$ are given by
\begin{eqnarray}
T^{ISW}_\ell = \int_{\eta_r}^{\eta_0} d\eta e^{-\tau(\eta)} \ j_\ell(k[\eta-\eta_0]) 
(c_{\Phi \Psi} \dot{\phi} - \dot{\psi}) \ , \\ 
M_\ell = c_{\delta \Psi} \int_{\eta_r}^{\eta_0} d\eta \ j_\ell(k[\eta-\eta_0])
\dot{z} W_g(z(\eta)) {\tilde \delta}(k,\eta)
\ ,
\end{eqnarray}
where the dot denotes differentiation with respect to conformal time $\eta$, 
$\eta_0$ is the time today, $\eta_r$ is a time very early in the
radiation era, $\tau(\eta)$ is the opaqueness, $W_g(z)$ is the normalized galaxy 
selection function,
$\dot{\phi}(k,\eta)$, $\dot{\psi}(k,\eta)$ and $\tilde{\delta}(k,\eta)$ are 
evolution functions which we define in the Appendix and which
can be calculated numerically using CMBFAST \cite{cmbfast},
$c_{\delta \Psi}$ and $c_{\Phi \Psi}$ are numerical coefficients also defined
in the Appendix, and $j_l(\cdot )$ are spherical Bessel functions.

The choice of the selection function $W_g(z)$ depends on
which large scale structure data set one wants to consider. Depending on
the particular experiment, one also has to account for the possible bias 
between the distribution of the observed objects and that of the underlying dark
matter. Our results for $C^{MT}(\theta)$ for 
the $\Lambda$CDM model are consistent with those of 
\cite{nolta,fosalba,scranton}, when appropriate biases and selection
functions are used.

Eq~(\ref{crosscor}) differs from the analogous expressions in
\cite{nolta,fosalba,scranton,bean,turok96,peiris00,cooray}
as it uses the primordial
curvature power spectrum rather than today's matter power spectrum. 
This, as explained in the Appendix, allows
us to take a more complete account of the dark energy perturbations.

\begin{figure}[tbp]
\centering
\scalebox{0.4}{\includegraphics{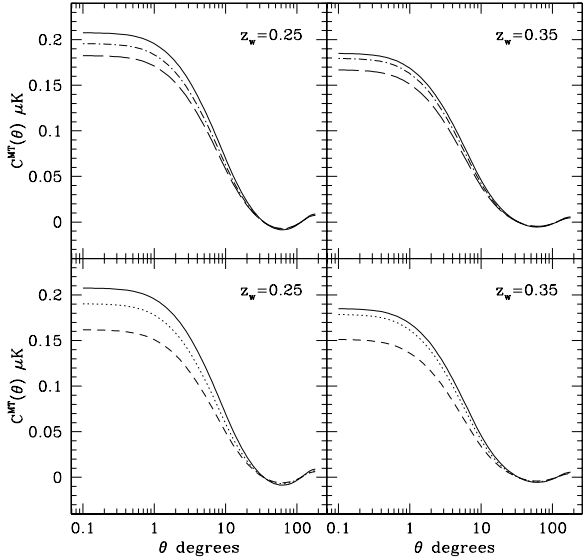}} 
\caption{The cross-correlation breaks the remaining degeneracies. 
In each panel, we plot a doomsday model together with its corresponding 
constant $w$ model (these two have degenerate CMB spectra and $P(k)$). 
For comparison, each plot also contains the fiducial $\Lambda$CDM model.  
Top two panels: $C^{MT}(\theta)$ for the $\tilde s=2$ (dot - short dash) 
and the $w=-0.81$ (long dash) models, as well as for the $\Lambda$CDM model 
(solid), for two values of $z_w$. Bottom two plots: the same as in the top 
two plots but for the $\tilde s=3$ (dot) and the $w=-0.66$ (short dash) models. }
\label{multiw} 
\end{figure}

\begin{figure}[tbp]
\centering
\scalebox{0.4}{\includegraphics{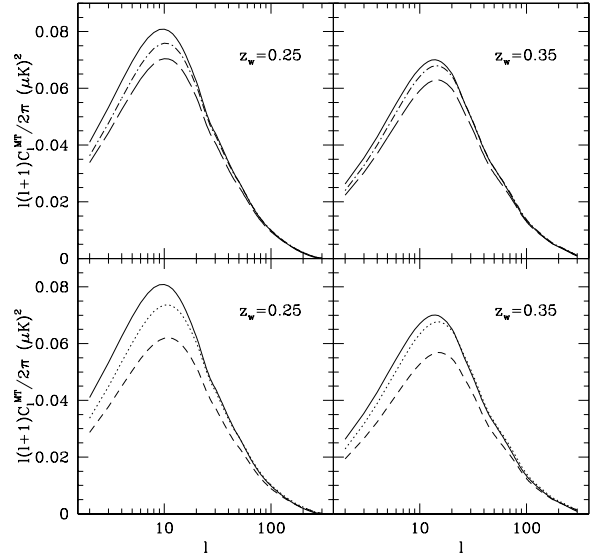}} 
\caption{Cross correlation angular spectra for the same models as in 
Fig.~\ref{multiw}.}
\label{multicl} 
\end{figure}

\begin{figure}[tbp]
\centering
\scalebox{0.4}{\includegraphics{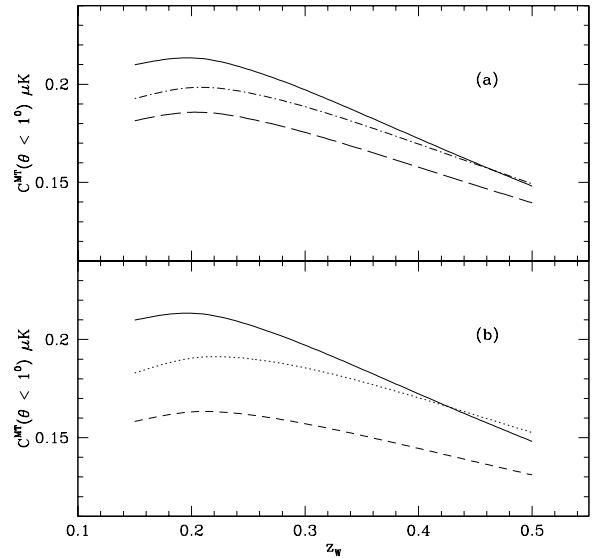}} 
\caption{$C^{MT}(0.1^\circ)$ vs $z_w$ for the models in Figs.~\ref{multiw}
and \ref{multicl}: $\Lambda$CDM (solid), $\tilde s=2$ (dot - short dash), 
$w=-0.81$ (long dash), $\tilde s=3$ (dot) and $w=-0.66$ (short dash).}
\label{ampl} 
\end{figure}

$C^{MT}(\theta)$ can be particularly useful for differentiating between
$w=const$ and varying $w$ models. The cross-correlation is sensitive
to the value of $w$ averaged over the range of the window function $W_g(z)$.
If $w$ is a rapidly changing function of redshift, as in the case of the
doomsday model, then depending on where the maximum of
the selection function is, $C^{MT}(\theta)$ will ``see''
different values of $\langle w \rangle$.
We have calculated the cross-correlation for several window functions,
all taken to be Gaussians of approximately the same width as the SDSS 
window functions \cite{scranton}, all with the same standard deviation 
$\sigma_w=0.07$
and centered
at various values of $z_w$ in the interval $[0.1,0.8]$. 
In Fig.~\ref{multiw} we show the plot of $C^{MT}(\theta)$ for the 
$\tilde s=2$ and $\tilde s=3$ models, together with their respective
$w={\rm constant}$ models,
at two different values of $z_w$. Corresponding angular spectra $C^{MT}_\ell$
are shown in Fig.~\ref{multicl}. In addition, in Fig.~\ref{ampl} we plot
the values of $C^{MT}(0.1^\circ)$ as a function of $z_w$. (As is seen
from Fig. \ref{multiw}, for angular
separations less than about a degree, the plot is insensitive to the choice 
of $\theta$.) From Fig.~\ref{ampl}
one can see that observations focusing on redshifts in the range  $z_w=[0.2,0.4]$
have the best potential of detecting the time-dependence of $w$.

\begin{figure}[tbp]
\centering
\scalebox{0.4}{\includegraphics{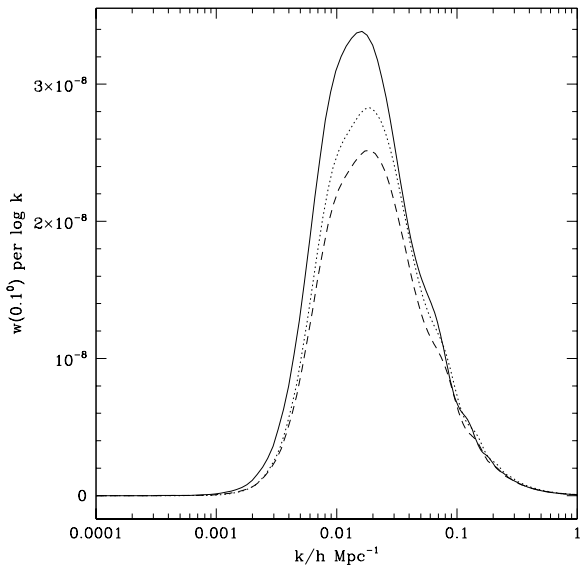}} 
\caption{$I(k)$ vs k for the $\Lambda$CDM (solid), $\tilde s=3$ (dot) 
and the $w=-0.66$ (short dash) models for $z_w=0.2$.}
\label{integrand} 
\end{figure}

An interesting question is which scales give the dominant contribution to the 
cross-correlation. In Fig.~\ref{integrand} we 
plot the contribution per log$(k)$ to the cross-correlation at $\theta=0.1^\circ$
for the $\Lambda$CDM, $\tilde s=3$ and $w=-0.66$ models using $z_w=0.2$. 
Namely, we define a quantity $I(k)$ by
\begin{equation}
C^{MT}(0.1^{o}) = \int d({\rm ln}k) \ I(k) \ .
\end{equation}
The expression of $I(k)$ can be deduced from Eq.~(\ref{wtg3}) of the Appendix.
As one can see from Fig.~\ref{integrand}, $I(k)$ has a broad peak  
around $k \sim .01 h$ Mpc$^{-1}$, corresponding to lengthscales in the 
range $20-300 h^{-1}$ Mpc. This roughly coincides with the peak in 
the matter power spectrum, depicted in Fig.~\ref{pvsk}. For larger angular 
scales, it is still the same linear scales that dominate the integrand. 
However, for angles $\theta \gtrsim 10^{o}$ there is destructive interference 
between the modes and the correlation gradually disappears.

Let us now briefly consider the prospects for observation, and whether these 
will allow us to distinguish amongst different models. In practice, one 
correlates the CMB anisotropies with some data set ({\it e.g} a galaxy survey) 
which is supposed to trace the underlying dark matter distribution. Of these two, 
the cleanest input is the CMB.
Although very precise, current CMB data cannot separate the ISW contribution 
from the net anisotropy, which reduces the signal-to-noise ratio
of the cross correlation. There is hope, however, that such separation 
may be possible in the future, with measurements of CMB polarization 
towards galaxy clusters \cite{CHB03}. 

A prime source of uncertainty in the matter distribution is the bias factor. 
One way to proceed (which has been implicitly used {\it e.g.} in \cite{fosalba}) 
is the following. From the observed CMB autocorrelation, and following the steps 
we have described in the previous Section, the matter power spectrum can be normalized. 
Comparing this to the autocorrelation function of any given matter survey, the 
bias factor can be inferred. Hence, from the cross correlation of the survey 
with the CMB data one can infer the cross correlation function of the matter 
distribution with the CMB. 
Note in particular that the predictions for the matter spectra for the $w=-0.66$ 
and $\tilde s=3$ models are practically indistinguishable (see. Fig.~\ref{pvsk}). 
Hence, adjusting for the bias will preserve the relative difference between 
predictions for the cross-correlation for these two models
\footnote{It should be noted that at small angles (where the cross-correlation with 
the matter power spectrum is maximal) and low redshifts, there is another 
contribution to the 
cross-correlation coming from the Sunjaev-Zeldovich (SZ) effect. Evidence for this 
contribution has already been reported in \cite{fosalba,afshordi}.
The SZ contribution 
depends on wavelength, and a number of future balloon experiments (as well as the 
Planck satellite) will be able to subtract this component out.} . 

Error bars in current determinations of the cross-correlation 
function \cite{xray,nolta,fosalba,scranton} are still too large 
to distinguish between models. However, the results in \cite{fosalba} 
have been obtained by using just a small fraction of the sky, and the 
situation may improve considerably with fuller sky coverage. Also, the 
uncertainties in bias may be substantially reduced with further observations 
such as weak lensing surveys. The differences in the cross correlation 
between the models considered in this Section can be as large as $20\%$, 
and they have a substantial dependence on redshift. Hence, it seems plausible 
that they may fall within the range of detectability in a not too distant future.

\section{Summary and conclusions}
\label{summary}

Current efforts to understand the dark energy component of our universe
often employ dynamical scalar fields. The dynamics of such scalar fields
lead to variation of the dark energy equation of state parameter, $w$. We have
considered astrophysical signatures of a varying equation of state, $w(z)$,
in the context of doomsday models. There are two main issues that we have
considered. First, can we tell if the average value of $w$ is different from 
$-1$. Secondly, what astrophysical signatures are sensitive to the time 
variation of $w$?

To address both issues, we have considered autocorrelations of the CMB temperature 
anisotropies, the matter power spectrum, and cross-correlation of the anisotropy
with matter fluctuations. The WMAP data is well reproduced by models with different
 average value of $w$ by suitably adjusting the Hubble parameter, and therefore 
 the CMB data cannot by itself discriminate amongst different models. For a given 
 CMB, the present linear matter power spectrum is sensitive to the average value 
 of $w(z)$, but not to its time dependence (see Fig. \ref{pvsk}). Finally, the 
 temperature-matter (TM)cross-correlation is sensitive to both the average value 
 of $w$ and to the time  variation (see Figs. \ref{multiw}, \ref{multicl} 
 and \ref{ampl}).

The TM cross-correlation requires accurate determination of both CMB anisotropies
and the matter fluctuations. Ongoing CMB observations have been very successful
at producing data with very small error bars and it is likely that the situation
will improve even further in the coming years. Accurate surveys of the matter 
fluctuations are likely to be more challenging since there are unknowns such
as the bias factor. However, since the CMB anisotropies and large-scale structure
originate from the same density fluctuations, it is quite possible that a 
combination
of the data can significantly reduce the uncertainties. We have shown that the 
optimal strategy 
for detecting a doomsday variation of $w(z)$ is to use a survey in the redshift 
range
$[0.2,0.4]$.

Observations of supernovae over the next ten years are also likely to
provide information on $w(z)$, as discussed in Ref. \cite{KKLLS03}.
The advantage of planned supernovae observations are that they provide
direct information on the Hubble expansion rate which is closely related
to $w(z)$. Furthermore, the data will be of good quality, with small error
bars. However, the observations may not be too sensitive to rapid variation of
$w(z)$ at recent redshifts, the hallmark of the doomsday scenario. 
The TM cross-correlations, on the other hand, are sensitive to this time
variation and may perhaps be used together with the supernovae observations 
to further constrain $w(z)$.

\acknowledgments
We would like to thank Robert Caldwell, Michael Doran, Enrique Gaztanaga, 
Dragan Huterer and Alex Vilenkin for very useful discussions
and, additionally, M.~Doran for cross-checking some of our preliminary
results with his CMBEASY code \cite{cmbeasy}.
The work of J.G. was supported by by CICYT Research Projects FPA2002-3598, 
FPA2002-00748, and DURSI 2001-SGR-0061. 
The work of T.V. was supported by DOE grant number DEFG0295ER40898 at CWRU.

\appendix

\section{}
\label{appendixA}

Here we explain details of our calculation of the cross-correlation.
Let us define
\begin{equation}
\Delta(\hat{\bf n}) \equiv {T(\hat{\bf n})- \bar{T} \over \bar{T}}
\end{equation}
and
\begin{equation}
\delta(\hat{\bf n}) \equiv 
{\rho(\hat{\bf n})- {\bar\rho} \over {\bar\rho}} \ ,
\end{equation}
where $T(\hat{\bf n})$ is the CMB temperature measured along the direction 
$\hat{\bf n}$,
$\rho(\hat{\bf n})$ is the mass density along $\hat{\bf n}$
\footnote{In reality one divides the sky into pixels, with a direction $\hat{\bf n}$
assigned to each pixel, and counts the number of galaxies, $N(\hat{\bf n})$ , 
inside each pixel. Then one can find the galaxy number overdensity inside each
pixel: $(N(\hat{\bf n})-{\bar N})/{\bar N}$, which would be related to 
$\rho(\hat{\bf n})$ up to a bias factor.}, and
$\bar{T}$ and $\bar{\rho}$ are the averaged CMB temperature and the
matter density.
The temperature anisotropy due to the ISW effect is an integral over the
conformal time:
\begin{equation}
\Delta(\hat{\bf n}) = \int_{\eta_r}^{\eta_0} d\eta \ e^{-\tau(\eta)}
( \dot{\Phi}-\dot{\Psi}) \left[(\eta_0-\eta) \hat{\bf n},\eta \right] ,
\end{equation}
where $\eta_r$ is some initial time deep in the radiation era, $\eta_0$
is the time today, $\Phi$ and $\Psi$ are the Newtonian gauge gravitational 
potentials\footnote{Throughout this Appendix we work in the Newtonian gauge
using conventions of, e.~g., Ref.~\cite{sasaki}.}, 
$\tau(\eta)$ is the opaqueness, which should, in principle, be included to 
account for the possibility of late reionization, 
and the dot denotes differentiation with respect to $\eta$.

The quantity $\delta(\hat{\bf n})$ contains contributions from astrophysical
objects (e.~g. galaxies) at different redshifts and can also be expressed as an 
integral over the conformal time:
\begin{equation}
\delta(\hat{\bf n}) = \int_{\eta_r}^{\eta_0} d\eta \ {dz \over d\eta} 
\ W_g(z(\eta)) \ 
{\delta}((\eta_0-\eta) \hat{\bf n},\eta) ,
\end{equation}
where $W_g(z)$ is a normalized galaxy selection function.

We are interested in calculating the cross-correlation function
\begin{equation}
C^{MT}(\theta) \equiv C^{MT}(|\hat{\bf n}_1-\hat{\bf n}_2|) \equiv 
\langle \Delta(\hat{\bf n}_1) \delta(\hat{\bf n}_2) \rangle \ ,
\end{equation}
where the angular brackets denote ensemble averaging and $\theta$ is
the angle between directions $\hat{\bf n}_1$ and $\hat{\bf n}_2$.
Let us introduce 
\begin{equation}
r \equiv \eta_0 - \eta \ .
\end{equation}
The Fourier decomposition for $\dot{\Phi}(r\hat{\bf n},\eta)$ can be
written as
\begin{equation}
\dot{\Phi}(r\hat{\bf n},\eta) = \int {d^3 {\bf k} \over (2\pi)^3} 
\dot{\Phi}({\bf k},\eta) e^{i{\bf k}\cdot \hat{\bf n}r} \ ,
\end{equation}
and similarly for $\dot{\Psi}(r\hat{\bf n},\eta)$ and 
$\delta_g(r\hat{\bf n},\eta)$.
We can write:
\begin{eqnarray}
C^{MT}(\theta) &=& 
\int_{\eta_r}^{\eta_0} d\eta_1 \int_{\eta_r}^{\eta_0} d\eta_2 \ 
\dot z(\eta_2) \ e^{-\tau(\eta_1)} W_g(z(\eta_2))  \nonumber \\ &\times&
\int {d^3 {\bf k} \over (2\pi)^3} \int {d^3 {\bf k'} \over (2\pi)^3} \ 
e^{i{\bf k}\cdot \hat{\bf n}_1 r_1} e^{i{\bf k}'\cdot \hat{\bf n}_2 r_2} 
\nonumber \\ &\times&
\langle [\dot{\Phi}({\bf k},\eta_1)-\dot{\Psi}({\bf k},\eta_1)]
\delta({\bf k'},\eta_2) \rangle  \ .
\label{wtg1}
\end{eqnarray}
Since the time-evolution of each Fourier mode only depends on the magnitude 
$k=|{\bf k}|$, we can separate the directional and time dependence as:
\begin{eqnarray}
\Phi({\bf k},\eta) &\equiv& \Phi({\bf k},\eta_r) \phi(k,\eta) \nonumber \\
\Psi({\bf k},\eta) &\equiv& \Psi({\bf k},\eta_r) \psi(k,\eta) \nonumber \\
\delta({\bf k},\eta) &\equiv& \delta({\bf k},\eta_r) \tilde{\delta}(k,\eta) \ .
\end{eqnarray}
Hence, we can write $\dot{\Phi}$ and $\dot{\Psi}$ as
\begin{eqnarray}
\dot{\Phi}({\bf k},\eta) &\equiv& \Phi({\bf k},\eta_r) \dot{\phi}(k,\eta) \nonumber \\
\dot{\Psi}({\bf k},\eta) &\equiv& \Psi({\bf k},\eta_r) \dot{\psi}(k,\eta) \nonumber \\
\end{eqnarray}
Consequently, the quantity $\langle [\dot{\Phi}({\bf k},\eta_1)-\dot{\Psi}({\bf k},\eta_1)] 
\delta({\bf k'},\eta_2) \rangle$ can be separated into the initial power 
spectra,
which contain all the information relevant to the ensemble averaging, and the 
time-evolving part which is the same for all members of the ensemble:
\begin{eqnarray}
\langle [\dot{\Phi}({\bf k},\eta_1)-\dot{\Psi}({\bf k},\eta_1)] 
\delta({\bf k'},\eta_2) \rangle &=& \nonumber \\
\langle \Phi({\bf k},\eta_r) \delta({\bf k'},\eta_r) \rangle
\dot{\phi}(k,\eta_1)\tilde{\delta}(k',\eta_2)  &-& \nonumber \\
\langle \Psi({\bf k},\eta_r) \delta({\bf k'},\eta_r) \rangle
\dot{\psi}(k,\eta_1)\tilde{\delta}(k',\eta_2)
\end{eqnarray}
We take $\eta_r$ to be a sufficiently early time in the radiation era when
all modes under consideration were superhorizon. Then, for adiabatic initial
conditions, the growing mode solutions for $\delta$, $\Phi$ and $\Psi$ are
related to each other via \cite{maber}:
\begin{equation}
c_{\delta \Psi} \equiv {\delta \over \Psi} = -{3\over 2} \ , 
\ c_{\Phi \Psi} \equiv {\Phi \over \Psi} = -\left(1+{2\over 5} R_\nu \right)  \ ,
\label{numconst}
\end{equation} 
where $R_\nu \equiv \rho_\nu / (\rho_\gamma +\rho_\nu)$ and $\rho_\nu$ is
the energy density in relativistic neutrinos. For $N_\nu$
flavors of relativistic neutrinos (we take $N_\nu=3$), after electron-positron 
pair annihilation, $\rho_\nu / \rho_\gamma = (7N_\nu/8)(4/11)^{4/3}$.
This allows us to write:
\begin{eqnarray}
&{}& \langle [\dot{\Phi}({\bf k},\eta_1)-\dot{\Psi}({\bf k},\eta_1)] 
\delta({\bf k'},\eta_2) \rangle = \nonumber \\ &=& c_{\delta \Psi}
\langle \Psi({\bf k},\eta_r) \Psi({\bf k'},\eta_r) \rangle \nonumber \\
&\times& \left[ c_{\Phi \Psi} \dot{\phi}(k,\eta_1) -
\dot{\psi}(k,\eta_1) \right] \tilde{\delta}(k',\eta_2)
\label{unif}
\end{eqnarray}
From homogeneity of space it follows that 
\begin{eqnarray}
\langle \Psi({\bf k},\eta_r) \Psi({\bf k'},\eta_r) \rangle =
(2\pi )^3 \delta^{(3)}({\bf k}+{\bf k'}) P_{\Psi}(k) \ ,
\label{homo}
\end{eqnarray}
where $P_\Psi(k)$ is the primordial gravitational power spectrum related to
the more frequently used curvature power spectrum 
$P_{\cal R}\equiv 2\pi^2 \Delta_{\cal R}^2  /k^3$ via 
\cite{Liddle,wmap_verde}
\begin{equation}
P_\Psi (k) = {9 \over 25} P_{\cal R} (k) 
= {9 \over 25} {2\pi^2 \over k^3} \Delta_{\cal R}^2 \ .
\label{psir}
\end{equation}

To the best of our knowledge, in all previous literature that contained
calculations of the cross-correlation, it was the matter power spectrum at
recent redshifts that was used, rather than the primordial spectrum. 
One can do that if fluctuations in the dark energy $\delta \rho_D$
are much smaller than those in cold dark matter, $\delta \rho_{cdm}$:
for $z \rightarrow 0$, one has
\begin{equation}
\Phi - \Psi \approx - {H^2 \over k^2}\left( \delta \rho_{cdm} + \delta \rho_D \right)
\end{equation}
and usually one proceeds by assuming that 
$\delta \rho_D \ll \delta \rho_{cdm}$ in the equation above.
While this is not necessarily an invalid condition,
working with the primordial spectrum allows us to use exact relations (\ref{numconst}),
valid deep in the radiation era, when all relevant modes are outside the horizon
and dark energy fluctuations are negligible,
and avoid the need for additional assumptions. It turns out, however, that,
while on scales $k \lesssim 0.001 h Mpc^{-1}$ and larger the fluctuations in
the dark energy can be as large as $10 \%$, on scales $k \sim .01 h Mpc^{-1}$, 
where the cross-correlation is important,
the contribution of dark energy perturbations is rather small, less than $1\%$.

Using Eqns.~(\ref{unif}) and (\ref{homo}) we can now re-write 
Eq.~(\ref{wtg1}) as
\begin{eqnarray}
C^{MT}(\theta) = {9 \over 25} \int_{\eta_r}^{\eta_0} d\eta_1 
\int_{\eta_r}^{\eta_0} d\eta_2 \ 
 \dot z(\eta_2) \ e^{-\tau(\eta_1)} W_g(z(\eta_2))  
\nonumber \\ \times  \int {d^3 {\bf k} \over 4\pi k^3} \Delta_{\cal R}^2(k) 
e^{i{\bf k}\cdot (\hat{\bf n}_1 r_1 - \hat{\bf n}_2 r_2)}
F(k,\eta_1,\eta_2) \ , \nonumber \\ {}
\label{wtg2}
\end{eqnarray}
where we have defined 
\begin{equation}
F(k,\eta_1,\eta_2) \equiv c_{\delta \Psi}
\left[ c_{\Phi \Psi} \dot{\phi}(k,\eta_1) -
\dot{\psi}(k,\eta_1) \right] \tilde{\delta}(k,\eta_2) \ .
\label{bigf}
\end{equation}
Decomposing the exponents in Eq.~(\ref{wtg2}) into spherical functions and 
some manipulations lead to:
\begin{eqnarray}
C^{MT}(\theta) &=& {9 \over 25} \int_{\eta_r}^{\eta_0} d\eta_1 
\int_{\eta_r}^{\eta_0} d\eta_2 \ 
\dot z(\eta_2) \ e^{-\tau(\eta_1)} W_g(z(\eta_2)) \nonumber \\ &\times&
\int {dk \over k} \Delta_{\cal R}^2(k) \ {{\rm sin}(kR)\over kR} 
\ F(k,\eta_1,\eta_2) \ ,
\nonumber \\ {}
\label{wtg3}
\end{eqnarray}
where $R \equiv \sqrt{r_1^2+r_2^2-2r_1r_2{\rm cos}\theta}$.
In addition, if one were to use the expression (\ref{wtg3}), one would have to
subtract the monopole and dipole contributions
to $C^{MT}(\theta)$. This can be achieved by taking
\begin{eqnarray}
\label{mondip}
&{}&{{\rm sin}(kR) \over kR} \rightarrow {{\rm sin}(kR) \over kR} 
- {{\rm sin} kr_1 \over kr_1}{{\rm sin} kr_2 \over kr_2}
\\ &{}& - {3\over k^2 r_1 r_2} \left({{\rm sin} kr_1 \over kr_1}-{\rm cos} kr_1 \right)
\left({{\rm sin} kr_2 \over kr_2}-{\rm cos} kr_2 \right) \nonumber
\end{eqnarray}

In practice, one wants to avoid evaluating double time integrals in Eq.~(\ref{wtg3}).
A common way to reduce them to a single time integration is to use the so-called  
small angle ($\theta \ll 1$) and small separation ($|r_1-r_2|\ll r_1$) 
approximations \cite{Limber}. These approximations were used in. e.~g. 
Refs.~\cite{fosalba,scranton}.
One can change the integration variables to $x=r_1-r_2$ and $r=(r_1+r_2)/2$
(or, equivalently, to $\eta=(\eta_1+\eta_2)/2$) and write, in this approximation:
\begin{eqnarray}
C^{MT}(\theta) &\approx& {9 \over 25} \int_{\eta_r}^{\eta_0} d\eta \ 
\dot z e^{-\tau(\eta)} \ W_g(z(\eta)) \nonumber 
\\ &\times& \int {dk \over k} \ 
\Delta_{\cal R}^2(k) \ F(k,\eta,\eta) \ \int_{-2r}^{2r} dx  \ {{\rm sin}(kR)\over kR} \ , 
\nonumber \\ {}
\label{wtg4}
\end{eqnarray}
where $R\approx \sqrt{x^2+r^2\theta^2}$.
The integral over $x$ can be evaluated analytically if one was allowed to 
replace the $[-2r,2r]$ limits by $[-\infty,\infty]$. One can assume that 
$2r$ is sufficiently large ($r > 1/k\theta$) on relevant scales for that 
replacement to be appropriate and use
\begin{equation}
\int_{-\infty}^{\infty} dx  \ {{\rm sin}(kR)\over kR} = {\pi \over k} J_0(kr\theta)
\end{equation}
to obtain the following form:
\begin{eqnarray}
C^{MT}(\theta) &\approx& {9 \over 25} \int_{\eta_r}^{\eta_0} d\eta \ 
 \dot z e^{-\tau(\eta)} \ W_g(z(\eta)) \int {\pi dk \over k^2} \nonumber \\ &\times&
\Delta_{\cal R}^2(k) \ J_0(k\theta[\eta_0-\eta]) \ F(k,\eta,\eta) \ .\nonumber \\ {}
\label{wtg5}
\end{eqnarray}
While the approximate expression (\ref{wtg5}) can be useful for analytical
estimates, it gives an error of order $2-4\%$ on the scales of interest
and we shall not resort to it.

Instead of evaluating the expression (\ref{wtg3}) directly, from computational
point of view, it is advantageous to decompose it into Legendre series, compute
the individual coefficients of the decomposition, and then sum the series.
Namely, Eq.~(\ref{wtg3}) can be written as
\begin{eqnarray}
C^{MT}(\theta)= 
\sum_{l=2}^{\infty} {2\ell+1 \over 4\pi} C^{MT}_\ell P_\ell({\rm cos} \ \theta) \ ,
\end{eqnarray}
where we do not include the monopole and dipole terms in the sum, and
where  $C^{MT}_\ell$ can be written as
\begin{equation}
C^{MT}_\ell = 4\pi {9 \over 25}\int {dk \over k} \ \Delta^2_{\cal R} \ 
T^{ISW}_\ell(k) \ M_\ell(k) \ ,
\label{crosscor1}
\end{equation}
with functions $T^{ISW}_\ell(k)$ and $M_\ell(k)$ defined as
\begin{eqnarray}
T^{ISW}_\ell = \int_{\eta_r}^{\eta_0} e^{-\tau(\eta)}d\eta \ j_\ell(k[\eta-\eta_0]) 
(c_{\Phi \Psi} \dot{\phi} - \dot{\psi})  \\ 
M_\ell = c_{\delta \Psi} \int_{\eta_r}^{\eta_0} d\eta \ j_\ell(k[\eta-\eta_0])
\dot{z} W_g(z(\eta)) {\tilde \delta}(k,\eta) \ ,
\end{eqnarray}
where $j_l(\cdot )$ are spherical Bessel functions.
One can use CMBFAST \cite{cmbfast}, with minor modifications, to compute 
functions $T^{ISW}_\ell$ and $M_\ell$ and to normalize $\Delta_{\cal R}^2(k)$.


\begin{thebibliography}{99}
\bibitem{mistake} This possibility was
pointed out in \cite{GV03}. However, an error in the analysis of the prior 
distribution after inflation lead to the conclusion that it always favors 
small values of $s$, implying that the equation of state
$w=-1$ should be expected in general. This has been corrected in 
Ref. \cite{GLV03} where it is shown that both situations are
possible, depending on the shape of the multifield potential $V(\phi_a)$.   
In particular, there is a wide class of models with prior
distributions favoring large $s$.
\end{thebibliography}
\end{document}